\documentclass[aip,cp,amsmath,amssymb,reprint]{revtex4-2}

\usepackage{graphicx}
\usepackage{dcolumn}
\usepackage{bm}
\usepackage[utf8]{inputenc}
\usepackage[T1]{fontenc}
\usepackage{mathptmx}
\usepackage{xspace}
\usepackage{hyperref}

\newcommand{\delight}{\textsc{delight}\xspace}

\newcommand{\annz}{\textsc{annz}\xspace}

\newcommand{\photoz}{photo-$z$\xspace}

\newcommand{\specz}{spec-$z$\xspace}

\begin{document}

%%%%%%%%%%%%%
%%% TITLE %%%
%%%%%%%%%%%%%
\title{Machine Learning Applications in Astrophysics: Photometric Redshift Estimation}

%%%%%%%%%%%%%%%%%%%%%%%%%%%%%
%%% AUTHOR AND AFFLIATION %%%
%%%%%%%%%%%%%%%%%%%%%%%%%%%%%
\author{John Y. H. Soo}
\email[Corresponding author: ]{johnsooyh@usm.my}
\author{Ishaq Yahya Khalfan Al Shuaili}
\author{Imdad Mahmud Pathi}
\affiliation{School of Physics, Universiti Sains Malaysia, 11800 USM, Pulau Pinang, Malaysia.}

\date{\today}

\begin{abstract}
Machine learning has rose to become an important research tool in the past decade, its application has been expanded to almost if not all disciplines known to mankind. Particularly, the use of machine learning in astrophysics research had a humble beginning in the early 1980s, it has rose and become widely used in many sub-fields today, driven by the vast availability of free astronomical data online. In this short review, we narrow our discussion to a single topic in astrophysics -- the estimation of photometric redshifts of galaxies and quasars, where we discuss its background, significance, and how machine learning has been used to improve its estimation methods in the past 20 years. We also show examples of some recent machine learning photometric redshift work done in Malaysia, affirming that machine learning is a viable and easy way a developing nation can contribute towards general research in astronomy and astrophysics.
\end{abstract}

\maketitle

%%%%%%%%%%%%%%%%%%%%%
%%% INTRODUCTION %%%%
%%%%%%%%%%%%%%%%%%%%%
\section{Introduction} \label{sec:intro}
% Can use \subsection{} and \subsubsection{}.
% Citations: use \cite{} or \onlinecite{}.
% \begin{eqnarray} for multiple numbered equations.
% Use \text{} for Roman text within a math environment.
% Use \begin{subequations} \begin{equation} for equations 1a, 1b etc.
The words 'artificial intelligence' and 'machine learning' have been the buzzword of the past decade, industries and governmental policies have been revolving around the development of tools related to them. While artificial intelligence is the attempt to create machines with human-like cognitive functions to think and solve problems, machine learning is merely a subset of it: it is a framework where machines can learn from data and find patterns between inputs and outputs provided and guided by humans. Machine learning has been used to solve various complex classification, regression, clustering, object detection and segmentation problems, speeding up processes which the limited human brain can do.

The viability of machine learning really depends on the availability of data: the more data one has, the better a machine will be able to learn. In the last century, astrophysics research has moved progressively from being theory-driven, observation-driven, and now data-driven \cite{longo_data_2014}. Currently with the existence of many large astronomical sky surveys, satellites and telescopes, astronomy has generated plenty of stellar and galaxy data, most of which has been processed and analysed, and remain openly accessible to the world. An example being the Strasbourg Astronomical Data Centre (CDS), which hosts the SIMBAD and VIZIER astronomical databases, providing physical data of millions of nearby stars and galaxies, as well as more than 21 000 deep-sky object catalogues \cite{ochsenbein_vizier_2000,wenger_simbad_2000}. Not to mention upcoming large sky surveys, like the Legacy Survey of Space and Time (LSST) and the Square Kilometre Array (SKA), which would provide us with data of unprecedented volume and quality, requiring state-of-the-art data storage and analysis peripheral to handle them \cite{ivezic_lsst:_2008,dewdney_square_2009}.

\begin{figure}
    \centering
    \includegraphics[width=0.65\linewidth]{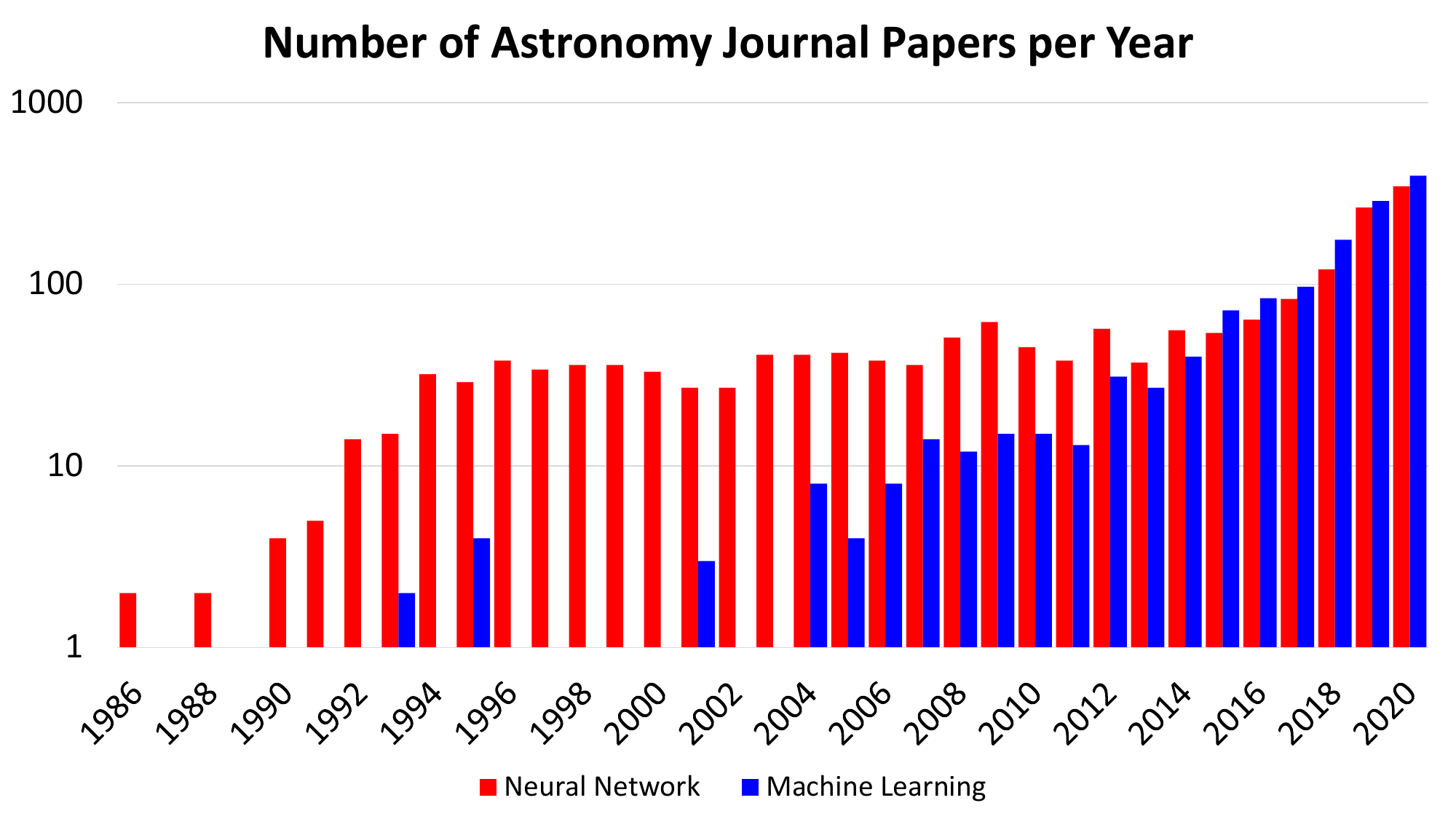}
    \caption{The number of astronomy and astrophysics journal papers with keywords 'neural network' or 'machine learning' increases with year. This figure is generated using data obtained from the \href{https://ui.adsabs.harvard.edu/}{NASA Astrophysical Data System (ADS)}.}\label{fig:journal} 
\end{figure}

Machine learning applications in astrophysics have been present since the late 1980s, as shown in Fig.~\ref{fig:journal}. The first instance where the word 'neural network' was used in an astronomy-related refereed journal paper was in 1986, when neural networks and simulated annealing where compared as optimisation methods for remote sensing data \cite{jeffrey_optimization_1986}. A review paper was later written in 1993 \cite{miller_review_1993}, suggesting that machine learning could be applied in telescopic adaptive optics, object classification and object detection.

During the 1990s, several prominent uses of machine learning in astrophysics include galaxy morphology classification \cite{storrie-lombardi_morphological_1992}, star-galaxy separation \cite{odewahn_star-galaxy_1993}, and stellar spectral classification \cite{gulati_stellar_1994}. Since then, the number of astronomy journal papers with keywords 'neural network' and 'machine learning' in their abstracts has increased, moving on an exponential trend since 2013. This is mainly due to the help of powerful and affordable computers, easier accessibility to machine learning algorithms and immense astrophysical data available online. 

After three decades since its introduction, machine learning remains a very useful tool in many areas of modern astronomy and astrophysics. Galaxy morphology classification remains an important application of machine learning \cite{banerji_galaxy_2010,zhang_classification_2021}, while solar flare and gravitational wave prediction and detection have benefited from machine learning analysis as well \cite{soni_discovering_2021,tang_solar_2021}. 

In this review, we narrow our machine-learning discussion to an astrophysics topic that has not been mentioned yet, i.e., the estimation of photometric redshifts of galaxies and quasars. In this work, we will discuss the background and significance of photometric redshifts, how machine learning has been used to improve its estimation methods in the past 20 years, its current outlook, and some examples of ongoing research in this field.

%%%%%%%%%%%%%%%%%%%%%%%%%%%%%
%%% PHOTOMETRIC REDSHIFTS %%%
%%%%%%%%%%%%%%%%%%%%%%%%%%%%%
\section{Photometric Redshifts}\label{sec:photoz}

Ever since the discovery of the expanding universe by Edwin Hubble \cite{hubble_relation_1929}, galaxy redshift has been used as a proxy to determine its distance away from the observer. Cosmological redshifts are obtained via measuring the shift in the galaxy spectra due to the expansion of space, and thus the higher the redshift, the further its distance from the observer. Physically, galaxy redshifts are obtained when their light pass through a dispersing element (e.g. a prism or grating), producing a spectra of a certain wavelength range, where signature emission or absorption lines within the spectra are then compared to elemental gas spectra on Earth. This kind of redshift is aptly known as spectroscopic redshift (\specz), which is expensive computationally, and also time-consuming.

As redshifts of billions of galaxies provide the essential third dimension to the 2D sky map we observe, spectroscopic methods to obtain redshifts are too slow and could not keep up to the current demand. Thus, photometric redshift (\photoz) methods have been used, where astronomers use broadband photometry to obtain redshifts of galaxies and quasars. In simple words, photometry can be regarded as the lower resolution version of spectroscopy, thus the derivation of photometric redshifts require some guessing work, finding theoretical or empirical relationships between broadband magnitudes of galaxies with their respective redshifts. Photometric redshifts can be easily generated in large amounts, and the focus of course is to improve and make them as accurate as possible within the limitations of equipment and statistical probabilities.

Photometric redshifts are widely sought after in astrophysics and cosmology research, in fact every large photometric survey produces them for various purposes with specific quality requirements. Among some of the most important uses of photometric redshifts include gravitational weak lensing analyses \cite{bonnett_using_2015,bilicki_bright_2021}, quasar studies \cite{curran_qso_2021}, galaxy clustering \cite{chan_z-sequence_2021} and even follow-ups for gravitational wave detection \cite{antolini_using_2016}. 

For full reviews of photometric redshifts, the reader is referred to \cite{brescia_data_2018,salvato_many_2019} for more information. In the following sections, we will give a brief discussion of the early history of photometric redshifts, how machine learning entered the picture and consequently left a great impacted in its research trajectory.

%%% EARLY DEVELOPMENT %%%
\subsection{Early Development}\label{sec:early}
Photometric redshifts started off as 'photoelectric red-shifts' in the 1960s, where Baum used a photometer and nine bandpasses to compare the displacement in spectral energy distribution between galaxies to predict redshifts of other galaxies \cite{baum_photoelectric_1962}. Early methods made use of the Lyman-drop technique and colour-colour diagrams to determine the relationship between magnitudes and redshifts \cite{partridge_._1974,koo_optical_1985}. Later on, the focus moved towards template fitting methods, where a predefined set of theoretical galaxy spectra are used to fit the photometry of galaxies, such that the photometric redshifts obtained are based on the best matched template at the best matched redshift. This method began with the work of Loh and Spillar \cite{loh_photometric_1986}, where many have later refined their templates for both elliptical and spiral galaxies, some theoretical and some empirical, in the following 20 years.

From the 2000s onwards, many different template fitting photometric redshift algorithms were developed and made publicly available, like \textsc{lephare} and \textsc{hyperz} \cite{arnouts_measuring_1999,bolzonella_photometric_2000}. Improvements were made in terms of algorithm as well as statistical characterisation, such as the inclusion of galaxy-type priors to constrain the redshifts obtained \cite{benitez_bayesian_2000}. At the same time, empirical methods were developed too, where simple polynomial empirical equations were constructed to relate redshift with the broadband magnitudes of galaxies. This method was originally pioneered in the 1990s \cite{connolly_slicing_1995}, and later improved and used to develop the colour-redshift relation for quasars \cite{richards_photometric_2001}.

%%% APPLICATION OF MACHINE LEARNING %%%
\subsection{Application of Machine Learning}\label{sec:machine}

In a certain way, the use of machine learning to estimate photometric redshifts can be thought of as a complex, highly non-linear and automated extension of the empirical methods implemented earlier. As the growth of computer technology took off exponentially in the late 1990s, machine-learning estimated photometric redshifts began to be explored. The first implementation of machine learning in photometric redshifts was probably done in 2003 \cite{firth_estimating_2003}, where artificial neural networks were used to obtain photometric redshifts for about 20~000 galaxies from the Sloan Digital Sky Survey (SDSS).

Since redshift is a continuous value, photometric redshift estimation is a regression problem, thus the determination of machine-learning photometric redshifts depends on a representative training set of galaxies which has both broadband photometry and spectroscopic redshifts. Following neural networks, different machine learning algorithms have been used to obtain photometric redshifts, e.g. support vector machines \cite{wadadekar_estimating_2005}, Gaussian processes \cite{way_novel_2006,almosallam_gpz:_2016}, k-nearest neighbours \cite{ball_robust_2007}, boosted decision trees \cite{gerdes_arborz:_2010}, self-organising maps \cite{way_can_2012,razim_improving_2021}, and many more.

As different machine-learning based photometric redshift algorithms have been created, many early efforts focused on improving existing algorithms \cite{sadeh_annz2:_2016} and comparing the performances between the different algorithms \cite{,hildebrandt_phat:_2010,bonnett_redshift_2016,schmidt_evaluation_2020}. This is significantly important in the recent years, as many current and upcoming sky surveys like the Dark Energy Survey (DES, \cite{abbott_dark_2005}), Legacy Survey of Time and Space (LSST, \cite{ivezic_lsst:_2008}) and the Physics of the Accelerating Survey (PAUS, \cite{padilla_physics_2019}) have stringent photometric redshift requirements to achieve, and efforts in finding the synergy between different photometric redshift algorithms play a crucial role in reaching the expected target.

Despite the overwhelming response towards machine learning within then photometric redshift community in the recent years, we note that template-based photometric redshift estimation methods are still important today. While template-based methods heavily depend on the choice of templates, filter responses and calibrations, machine-learning methods heavily depend on the size and representation of the training set for accurate results. Template-based methods are more physical, they link back to the theory and model of galaxies; while machine-learning methods let the data speak for themselves, and even allow non-photometric data (e.g. galaxy size, ellipticity) be used to improve redshift accuracy \cite{soo_morpho-z:_2018,menou_morpho-photometric_2019}. 

\begin{figure} %%% FIGURE: TEMP vs ML %%%
    \centering
    \includegraphics[width=0.8\linewidth]{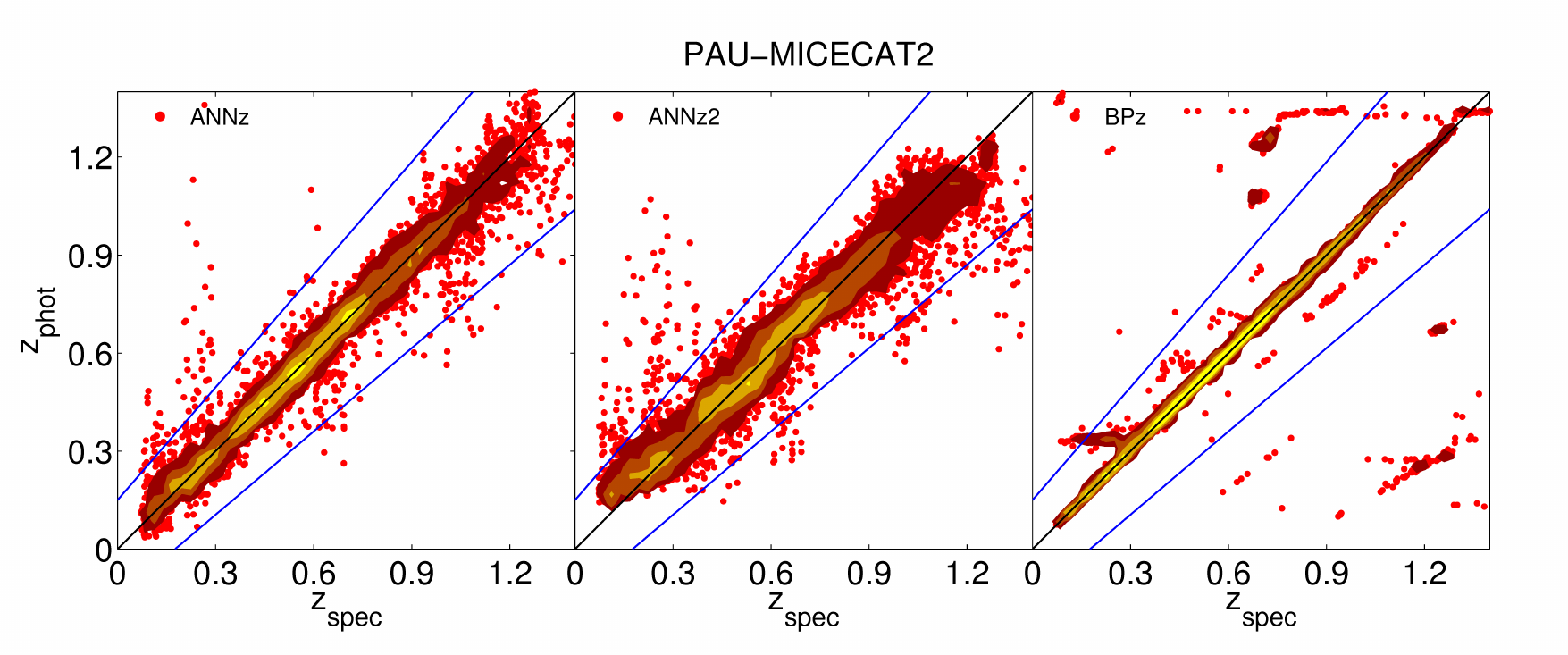}
    \caption{Plots of photometric redshift ($z_\textrm{phot}$) vs. spectroscopic redshift ($z_\textrm{spec}$) of galaxies in the PAU-MICECAT2 simulated catalogue \cite{carretero_cosmohub_2018}, comparing two machine-learning algorithms \textsc{annz} \cite{collister_annz:_2004} and \textsc{annz2} \cite{sadeh_annz2:_2016} with one template-based algorithm \textsc{bpz} \cite{benitez_bayesian_2000}. The plots show that machine-learning algorithms produced less scatter, while the template-based method produced a tighter relationship. This catalogues uses 6$ugrizy$ broad bands and 40 narrow bands similar to PAUS \cite{marti_precise_2014}.}\label{fig:micecat} 
\end{figure}

Performance-wise, machine learning methods produce lower scatter, however template-based methods produce a tighter relationship between the photometric and spectroscopic redshifts, as shown in Fig.~\ref{fig:micecat}. Thus in order to obtain the best of both worlds, the synergy between both methods should be explored to further improve photometric redshift results. An example being the use of Gaussian processes to model better galaxy templates to fit to the data, creating a hybrid method to estimate photometric redshifts \cite{soo_pau_2021}.

%%% CURRENT OUTLOOK %%%
\subsection{Current Outlook}\label{sec:outlook}

Recent work on photometric redshifts focus on the statistical representation of photometric redshifts, ensuring that their error derivation and probability distribution function (PDF) are good representations of the photometric redshift derived. One example is the use of k-nearest neighbours to determine the photometric redshift error, which was found to be a better method than traditional propagation of errors \cite{sadeh_annz2:_2016}. Other related work include the use of self-organising maps to detect anomalies \cite{razim_improving_2021}; recursive training to improve the output \cite{zitlau_stacking_2016}; and feature selection techniques to optimise the number of inputs used \cite{disanto_return_2018}.

With the efficiency of computers improving in the past few years, deep learning and convoluted neural networks have also been used to determine photometric redshifts straight from galaxy images \cite{disanto_photometric_2018}. In a recent work, transfer learning has been used to develop the deep learning code \textsc{deepz} in PAUS, where it is found that the scatter statistic is reduced by 50 per cent as compared to those produced by existing algorithms \cite{eriksen_pau_2020}.

Therefore, it is definitely promising that deep learning applications in photometric redshifts would remain essential in the coming years. For more information, the reader is referred to \cite{feigelson_twenty-first-century_2021} for a review on recent advancements in photometric redshift research.

%%%%%%%%%%%%%%%%%%%%
%%% ONGOING WORK %%%
%%%%%%%%%%%%%%%%%%%%
\section{Ongoing Work}\label{sec:work}

In the following sections, we briefly discuss several ongoing work which aimed to improve photometric redshift estimation using various machine learning methods.

%%% SYNERGY BETWEEN TEMPLATE AND ML %%%
\subsection{Synergy of Template and Machine Learning Methods}\label{sec:synergy}

In an effort to test the synergy between machine learning and template-based photometric redshift algorithms, we used \textsc{annz2} \cite{sadeh_annz2:_2016} and \textsc{bpz} \cite{benitez_bayesian_2000} to produce photometric redshift PDFs for galaxies from the Luminous Red Galaxy (LRG) sample \cite{eisenstein_spectroscopic_2001} and the Stripe-82 Coadd sample \cite{annis_sloan_2014}, trained using spectroscopic redshifts from SDSS \cite{york_sloan_2000}. We attempted several ways of combining the PDF distributions to see if we could get better photometric redshifts out of both methods. Early results have shown that in both samples, a simple averaging of the PDFs produced by both algorithms do not yield better results than both algorithms (Al Shuaili et al., in prep.). Further tests are still being conducted to see if weights can be used to better improve the results.

%%% ACTIVATION FUNCTION %%%
\subsection{Activation Functions in ANNz}\label{sec:activation}

The machine-learning photometric redshift algorithm \textsc{annz} \cite{collister_annz:_2004} has been around for almost two decades, and it has been used to determine photometric redshifts for various galaxy samples \cite{collister_megaz-lrg:_2007,abdalla_comparison_2011}. As the architecture of neural networks have improved through the years, we attempt to assess the viability of this C++ code today by changing and its original activation function (sigmoid) to other more modern activation functions widely used in recent literature (see \cite{szandala_review_2021} for a compilation and discussion of these activation functions).

\begin{figure}
    \centering
    \includegraphics[width=0.4\linewidth]{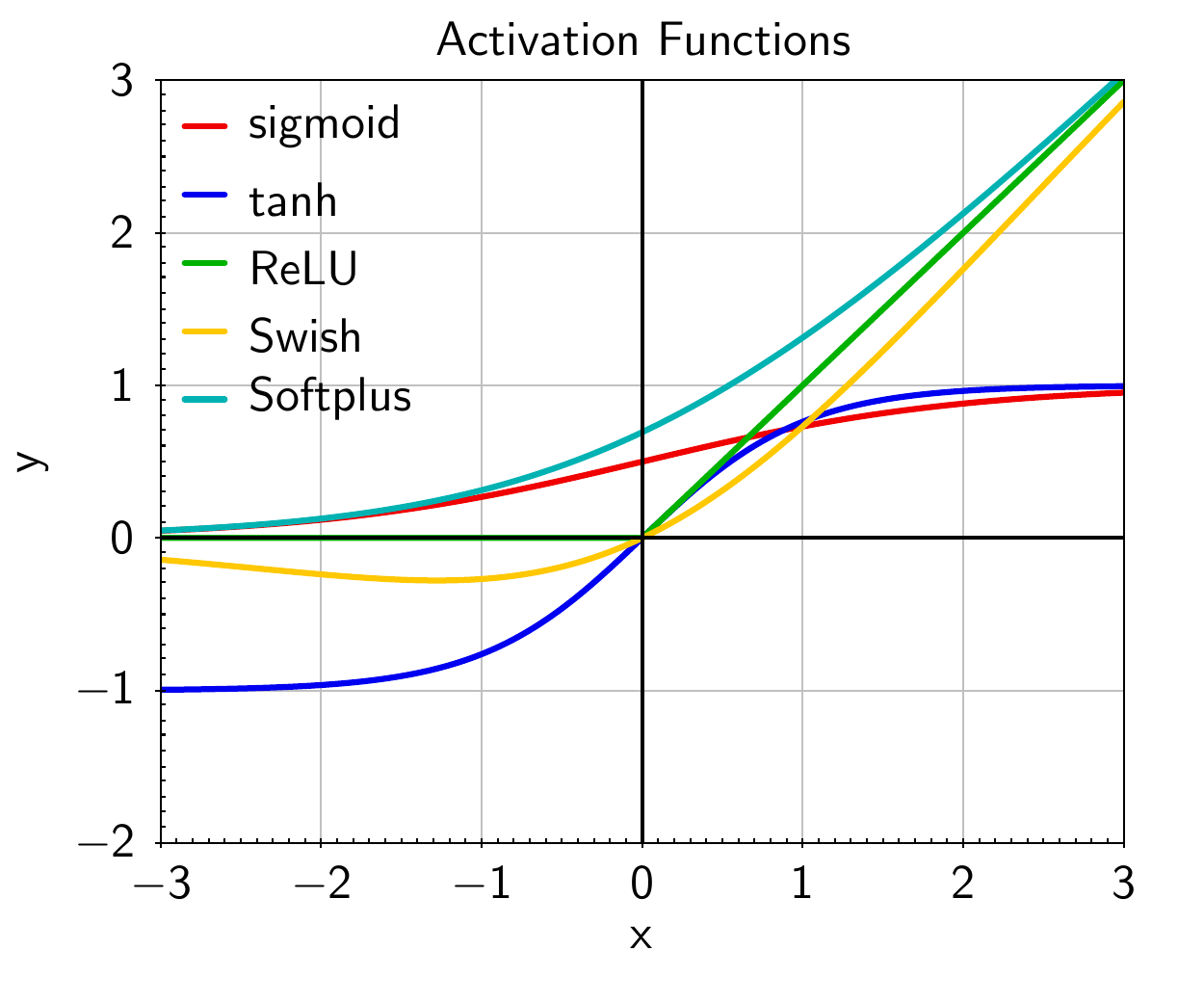}
    \caption{The visualisation of the activation functions sigmoid, tanh, ReLU, swish and softplus on the $x$-$y$ plane, used in Mahmud Pathi et al. (in prep.).}\label{fig:activationtype} 
\end{figure}

Using the same LRG and Stripe-82 galaxy samples, we modify the original C++ library of \textsc{annz} code to include the activation functions tanh, swish, softplus and rectified linear unit (ReLU). These activation functions are used to add non-linearity into the feed-forward and back-propagation process, and particularly the tanh and ReLU functions are widely used in deep learning processes \cite{szandala_review_2021}. The equations of the activation functions mentioned are visualised in Fig.~\ref{fig:activationtype}.

\begin{figure}
    \centering
    \includegraphics[width=\linewidth]{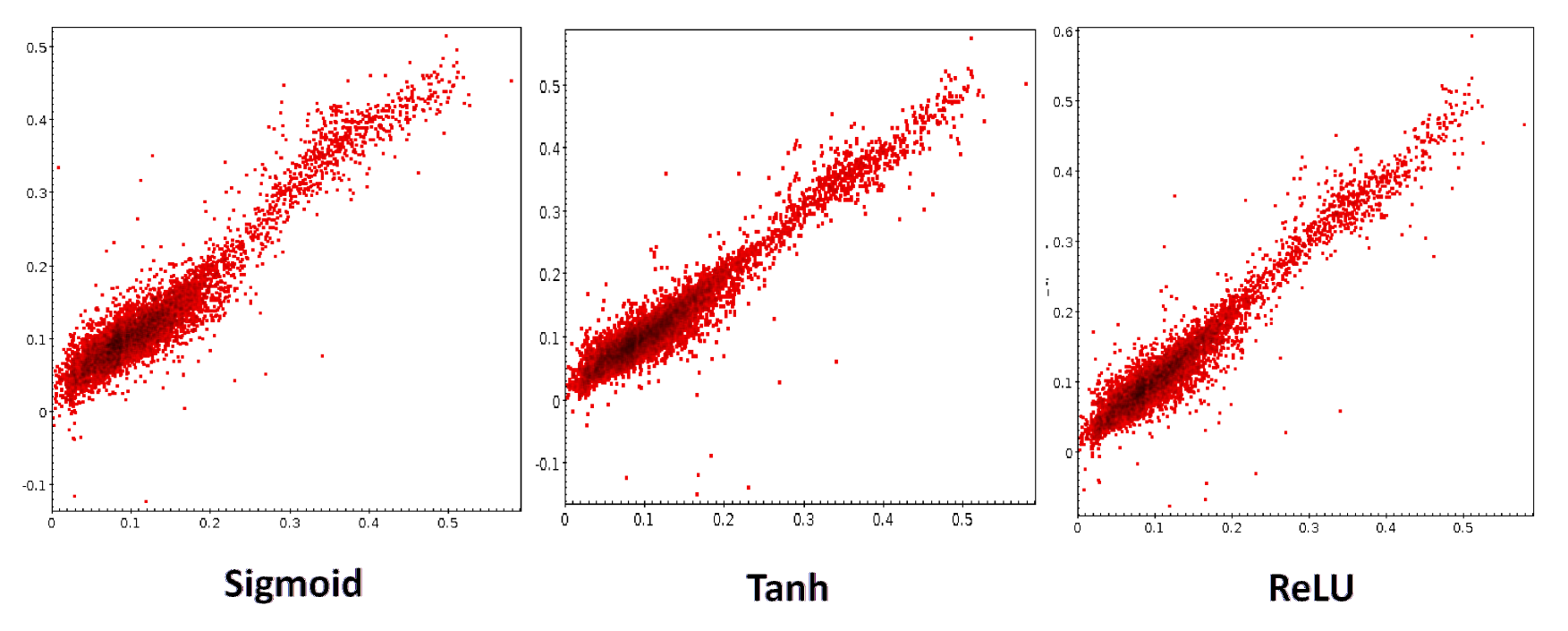}
    \caption{Plots of photometric redshift ($z_\textrm{phot}$) vs. spectroscopic redshift ($z_\textrm{spec}$) of Stripe-82 galaxies, comparing the results when different activation functions are used. Early results show that the ReLU activation function improves the root-mean-square error by 13.8 per cent.}\label{fig:activation} 
\end{figure}

Preliminary results have shown that while the improvement is not large on the LRG sample, significant improvement is found on the Stripe-82 sample, which contains fainter galaxies. Particularly, the ReLU activation function gives an improvement of 13.8 per cent in root-mean-square error, reaching a value of $\sigma_\textrm{RMS}=0.02168$ (Mahmud Pathi et al., in prep.). The performance of photometric redshifts produced for the Stripe-82 sample using different activation functions are shown in Fig.~\ref{fig:activation}.

We have also found that the performance of sigmoid and tanh degrades while ReLU improves as the network architecture gets more complex (i.e. higher number of hidden layers and nodes). Further fine-tuning is currently being conducted to search for optimum architectures for each activation function used, and finally conclude if \annz, an algorithm designed about 20 years ago, remains as competitive as other photometric redshift algorithms designed in the recent years.

%%% PAUS-GALFORM %%%
\subsection{Photometric Redshifts for the PAUS-GALFORM Simulated Sample}\label{sec:galform}

The Physics of the Accelerating Universe Survey (PAUS, \cite{padilla_physics_2019}) is a photometric sky survey which observes the sky using 40 narrow bands, tasked to map the large scale structure of the universe up to $i\sim 25$. With the help of narrow bands of widths about 100 \r{A}, it is expected that the quality of photometric redshifts would reach a root-mean-square error precision of $\sigma_\textrm{RMS}<0.0035(1+z)$. 

In an upcoming collaborative work, attempts have been made to calibrate and compare the performance of three photometric redshift algorithms, namely \textsc{delight} \cite{leistedt_data-driven_2017}, \textsc{cigale} \cite{boquien_cigale_2019} and \textsc{bcnz2} \cite{eriksen_pau_2019} on the PAUS-GALFORM simulated galaxy sample. This mock galaxy catalogue is a subsample of a lightcone generated using \textsc{galform} \cite{cole_hierarchical_2000}, a semi-analytical model of galaxy formation. The mock galaxy catalogue has galaxy fluxes generated to mimic that of the photometry of PAUS, added with realistic noises and flux errors.

\begin{figure}
    \centering
    \includegraphics[width=0.4\linewidth,page=1]{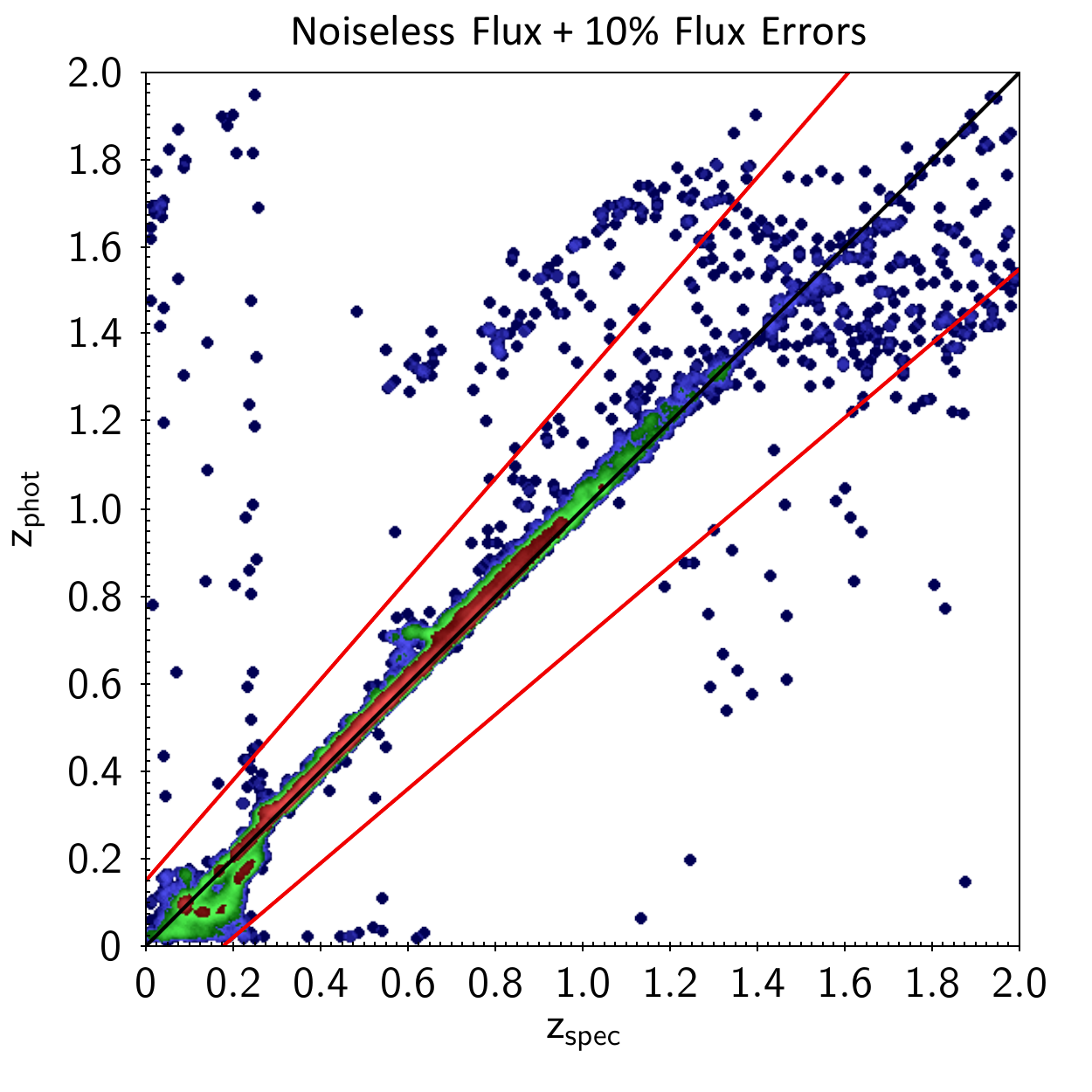}
    \includegraphics[width=0.4\linewidth,page=2]{figure4.pdf}
    \caption{Plots of photometric redshift ($z_\textrm{phot}$) produced by \textsc{delight} vs. spectroscopic redshift ($z_\textrm{spec}$) for the PAUS-GALFORM mock galaxy sample. Different types of flux errors are tested in order to assess their effects on the photometric redshift quality.}\label{fig:pausgalform} 
\end{figure}

Early results using \textsc{delight} have shown that the photometric redshift 68th percentile error (see \cite{soo_pau_2021} for definition) doubles when real noise and errors are added to the fluxes. As shown in Fig.~\ref{fig:pausgalform}, the relationship is tighter before real noise is added, giving a scatter of $\sigma_\textrm{RMS}=0.0794$ and an outlier rate of 1.85 per cent (Soo et al., in prep.). Once real noise has been added, the relationship is less tight, however the catastrophic outliers are reduced instead. This work will contribute to better understanding of how real flux errors contribute to the photometric redshift uncertainty, and how it can be characterised. As \delight is a hybrid template-machine-learning algorithm, we look forward to compare this results with the other two template algorithms and observe any trends or anomalies within the sample.

%%% CONCLUSION AND FUTURE WORK %%%
\section{Conclusion and Future Work}

As tools of artificial intelligence continue to be popularised in the industry and academia, we are positive that machine learning would continue to be relevant in astrophysics research in the coming decade. We expect to see more applications of machine learning, particularly in deep learning, object detection and segmentation in astrophysics in the decades to come. Thus it is also important for researchers and educators to help equip current university students with data literacy skills in preparation for such research trajectories.

We would also like to point out that, machine learning is in fact a great opportunity for researchers based in developing countries like Malaysia. This is especially true in the field of astronomy, where access to large survey collaborations and telescopic facilities are region-limited and government-dependent. Simple machine learning algorithms can be run easily on flagship computers on the market, thus with easy access to free astronomical data available on the internet, many data-driven research questions can be created and answered. We encourage budding young researchers to embrace opportunities to learn about machine learning, as it will definitely come in handy in this era of big data we live in.

%%%%%%%%%%%%%%%%%%%%%%%
%%% ACKNOWLEDGEMENT %%%
%%%%%%%%%%%%%%%%%%%%%%%
\begin{acknowledgments}
JYHS expresses his gratitude towards the organisers of COMDATA2021 for the opportunity to give a contributed talk in this conference. JYHS and IMP acknowledge financial support from the Short Term Grant (Skim Geran Penyelidikan Jangka Pendek), Universiti Sains Malaysia with account number 304/PFIZIK/6315395. IMP also acknowledges financial assistance under the Gra-Assist 2020 scheme provided by the Institute of Postgraduate Studies (IPS), Universiti Sains Malaysia.
\end{acknowledgments}

%\nocite{*}
\bibliographystyle{apsrev4-2}
\bibliography{photoz_comdata}% Produces the bibliography via BibTeX.

\end{document}